# Coupling characterization and noise studies of the Optical Metrology System on-board the LISA Pathfinder Mission

G. Hechenblaikner[1*], R. Gerndt[1], U. Johann[1], P. Luetzow-Wentzky[1], V. Wand[1],

H. Audley[2], K. Danzmann[2], A. Garcia-Marin[2], G. Heinzel[2], M. Nofrarias[2], F. Steier[2]

[1] *EADS Astrium GmbH, Friedrichshafen, Germany*

[2] *Albert Einstein Institut, Hannover, Germany*

[*]*Corresponding author: Gerald.Hechenblaikner@astrium.eads.net*

In this article we describe the first investigations of the complete engineering model of the Optical Metrology System (OMS), a key subsystem of the LISA Pathfinder science mission to space. The latter itself is a technological precursor mission to LISA, a space-borne gravitational wave detector. At its core, the OMS consists of four heterodyne Mach Zehnder interferometers, a highly stable laser with external modulator and a phase-meter. It is designed to monitor and track the longitudinal motion and attitude of two floating test-masses in the optical reference frame with a (relative) precision in the picometer and nanorad range, respectively. We analyze sensor signal correlations and determine a physical sensor noise limit. The coupling parameters between motional degrees of freedom and interferometer signals are analytically derived and compared to measurements. We also measure adverse cross-coupling effects originating from system imperfections and limitations and describe algorithmic mitigation techniques to overcome some of them. Their impact on system performance is analyzed in the context of the Pathfinder mission.

# I. Introduction

The optical metrology system (OMS) [1, 2] represents a key part of the LISA Technology Package, the scientific payload for the LISA Pathfinder (LPF) mission, due to be launched in 2012 by the European Space Agency [3]. LISA Pathfinder, besides major other tasks, will demonstrate the operation of an interferometer with two freely geodetically floating test-masses in its path and will be the most precise geodesics explorer flown as of today. It aims to demonstrate the technological basis required to perform measurements of the residual test-mass acceleration $a_r$ better than $3 \times 10^{-14} \, ms^{-2} Hz^{-1/2}$, relaxing towards higher frequencies as given in Equation 1 for the linear spectral density (LSD) of $a_r$.

$$LSD(a_r) \leq 3 \times 10^{-14} \sqrt{1+\left(\frac{f}{3mHz}\right)^4} \, ms^{-2} Hz^{-1/2}, f \geq 1mHz \qquad (1)$$

LPF is essentially a technological precursor mission to LISA (Laser Interferometer Space Antenna), the actual mission to detect gravitational waves based on interferometry [4]. In LISA the beams will propagate between 3 spacecraft in a triangular constellation of 5 million km side-length. In Pathfinder the distance between the test-masses is shrunk to only 38 cm so that the arm-length is far too small to detect actual gravitational waves. Furthermore, the sensitivity requirements are somewhat relaxed compared to those of LISA, where technological improvements and lessons learnt will help to outclass the performance of LISA Pathfinder. Nonetheless, many of the measurement principles, key technologies, and underlying physical noise sources to be characterized and studied are similar in the two missions, which makes Pathfinder a crucial milestone, as indicated by its name, on the way towards successful LISA mission.

There are two important aspects of LISA Pathfinder which we emphasize at this point because they constitute the higher-level motivation for the type of measurements described in this article and also determine the accuracy and extent of our data analysis.

One aspect is the demonstration of two quasi freely floating test-masses as part of the "drag free attitude and control system" (DFACS) [5, 6]. Note that some degrees of freedom are removed by electrostatic suspension while motion along the axis between the two test-masses is left unconstrained. In "science mode" DFACS monitors the test-mass position relative to the electrodes integrated in the walls of their confining "electrode cage" and maneuvers the spacecraft using micro-Newton thrusters in such a way as to avoid any collision of the drifting test-mass with the electrode housing.

The other aspect is the optical metrology system, which essentially allows the precise measurement of test-mass position and attitude to provide DFACS with the required feedback to steer the spacecraft and test-masses accordingly. As an added feature, the OMS provides raw data to accurately determine the residual test-mass acceleration which constitutes a basis to study the noise environment occurring onboard the spacecraft. The correct alignment of the OMS reference frame with the spacecraft reference frame, in particular with the electrode housing frame, is essential for that goal [7]. Noise, alignment and cross-coupling are all crucial factors in that respect and therefore central topics of our investigations in this article.

We give a detailed account of how the optical metrology system is characterized and operated, describe the system parameters and their inter-dependencies, and establish the optimal operating points. A thorough understanding of the sensitive system features lays the foundations for optimal performance of the system and also sheds light onto its limitations and inaccuracies - with direct impact on mission performance and operations. We purposely leave out the laser stabilization loops and performance measurements, as this would be beyond the scope of this article, and refer to other dedicated articles on this topic [8, 9, 10].

The article is structured as follows:

**II. Basic operating principle**: The essential core structure of the Optical Metrology System is described in a level of detail that is required to understand the succeeding measurements and the implications of the measurement results.

**III. Phase shifts and channel noise:** In the first step we determine the phase-offsets between the processing channels (phase-meter, diode and software processing) and measure the density distributions of the phase-fluctuations. We then derive the noise figures of all channels and calculate the noise correlations between the individual channels, allowing us to derive a physical limit of $\sim 0.5\, pm/\sqrt{Hz}$ for the position noise in the absence of laser frequency fluctuations.

**IV. Interferometer coupling parameters:** In the next step we derive analytical expressions for the coupling parameters describing the relation between the test-mass orientation and the corresponding phase from the interference signals. Then the actual coupling parameters are measured and the used techniques and their limitations are described in detail.

**V. Cross couplings terms:** We examine the cross-coupling between various test-mass degrees of freedom and discuss the impact on measurement accuracy and system performance. A crucial parameter in the cross-coupling strength is the orientation of the sensor (quadrant diode) reference frame with respect to the nominal bench frame, which can be inferred from the measurement data. The data also contain information on the beam misalignments with respect to the diode centers, the measurement beam size and profile on the diodes, and the beam power ratios at certain reference points, all of which have a direct impact on coupling parameters. We then test the ability to monitor large test-mass displacements through continuous tracking of the change in longitudinal phase. At the same time the accuracy to which the observed motion of test-mass 1 decouples from the motion of test-mass 2 is determined.

**VI. Conclusions and outlook:** We conclude with a summary of the measurement results and point out the system limitations. The latter arise as a combination of manufacturing tolerances and hardware limitations on the one hand, and measurement errors and inaccuracies due to constraints in the testing environment on the other hand.

## II. Basic operating principle

We shall only briefly describe the basic operating principle of the Optical Metrology System and refer the reader to other publications for more detail [1, 2, 3]. For information on the basics of heterodyne interferometry, which is at the core of the OMS, we refer the reader to [11].

The OMS comprises the following units:

1. **The optical bench interferometer:** It consists of four heterodyne Mach-Zehnder interferometers, each equipped with two quadrant photo-diodes for interference detection. Note that the "test-masses" which are freely floating in space and reflect the "measurement" beam in two of the four interferometers are substituted for "dummy mirrors" in the test setup.

2. **The laser unit:** Its stable single-mode output of wave-length $\lambda = 1.064\,\mu m$ is split into "reference" and "measurement" beams which separately pass through an external **laser modulator**. The two modulator output beams are brought to interfere on the quadrant diodes of each interferometer. They are frequency-shifted by the heterodyne frequency $f_{het} = 1kHz$ relative to another which constitutes the primary beat frequency of the interference pattern.

3. **The phase-meter** samples the interference signals from the photo-diodes at $f_s = 50 kHz$, and applies a discrete Fourier transform (DFT) on a time series of length $T_{DFT} = 10 ms$. Only the complex amplitude of the frequency bin centered around $f_{het}$ and the real value of the zero frequency bin (DC) are retained and transmitted to the DMU at a rate of 100 $Hz$.

4. **The data management unit** (DMU) with the OMS application software receives the DFT-data from the phase-meter and continues processing them. It calculates longitudinal and differential phases and infers test-mass position and attitude from them. For reasons related to limitations in communication bandwidth, data are down-sampled from 100 Hz to 10 Hz by application of a moving average filter before they can be communicated from the DMU to the experimental test-facility where they are displayed, recorded and stored for later retrieval.

As the optical bench plays a central role and is rather sophisticated in its design, we shall highlight its basic functionality in the following paragraph. A schematic of the optical bench is given in Figure 1. It is comprised of four heterodyne Mach Zehnder interferometers, referred to as "x1", "x1-x2", "frequency" and "reference" interferometer with the frequently used suffix 1, 12, F, and R, respectively. Each interferometer is equipped with two quadrant photo-diodes, the nominal diode "A" and the redundant diode "B" after the recombination beam splitter; e.g. the interference pattern of interferometer "x1" is detected by photo-diodes PD1A and PD1B. The redundant interferometer arms and diodes are not further used and investigated in this article. A detailed description on their use and functionality is available in [12].

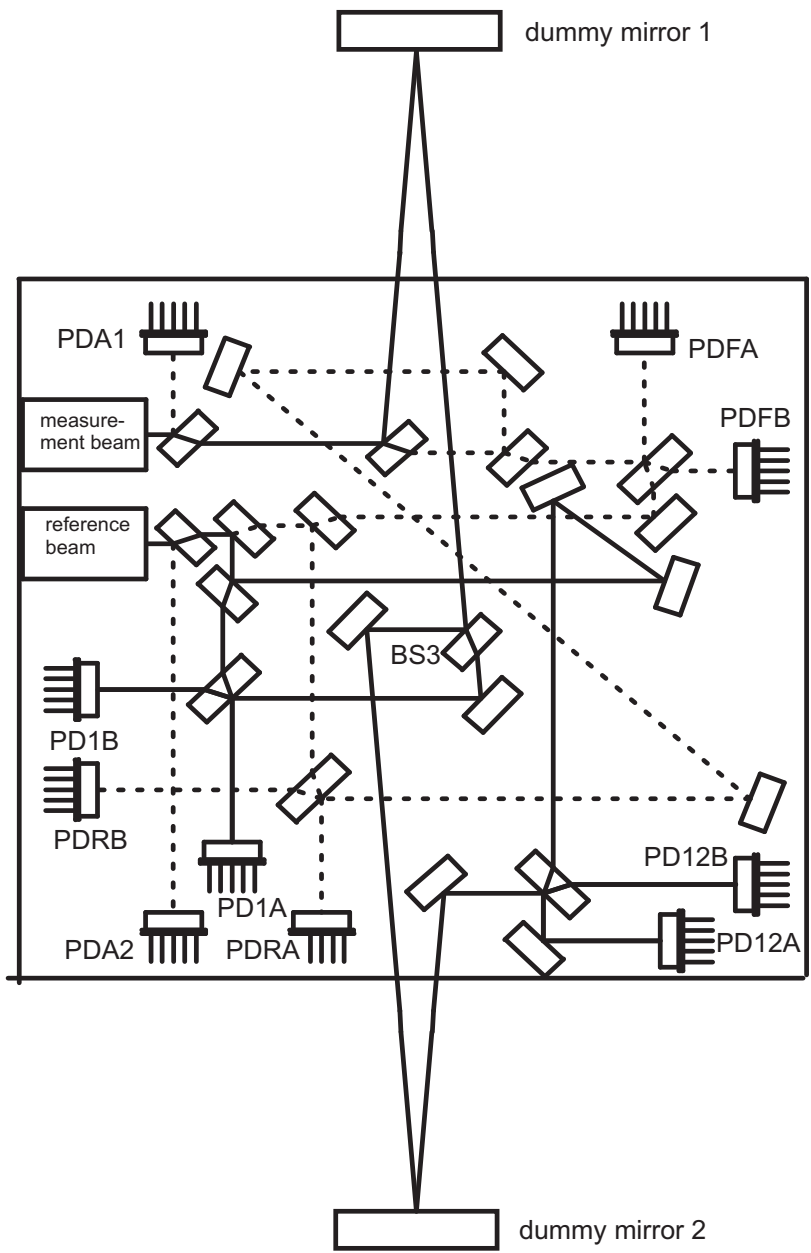

**Figure 1: Schematic of the optical bench (drawn to scale) comprised of four independent heterodyne Mach-Zehnder interferometers. The solid lines mark the beam paths in the x1 and the x1-x2 interferometers. The dotted lines mark the beam paths of the reference and frequency interferometers.**

Interferometer "x1" determines the position and attitude of test-mass 1 elative to the optical bench, interferometer "x1-x2" the relative position and attitude of test-mass 2 with respect to test-mass 1. The reference interferometer provides a reference phase $\Psi_R$ which is subtracted from the phases of all other interferometers through the processing software. This effectively cancels optical path-length variations which occur before the beams are split by the first optical elements on the bench, in particular phase variations from transmission through the optical fibres or the laser modulator are compensated. Note that the optical bench which is made of Zerodur provides inherently very low thermal expansion. It has no movable components and all silica mirrors and beam splitters are hydroxyl-catalysis bonded [13] such that the entire structure forms a quasi-monolithic entity with excellent mechanical and thermal properties.

Additionally the reference phase serves as error signal for a feedback loop to compensate the adversary effects of optical side-bands in the laser frequency spectrum which appear as a consequence of radio-frequency cross-talk inside the laser modulator [8, 9]. The frequency interferometer "F" translates laser frequency noise into phase-noise through a deliberate mismatch of the optical path length of the two interfering beams (the phase noise scales proportional to frequency fluctuations and optical path-length difference). The phase of the frequency interferometer serves as error signal to close two feedback loops for laser frequency stabilization ("fast" and "slow" loop), which actuate the laser cavity length through changes in mechanical stress and temperature, respectively. Laser power fluctuations are stabilized through a "fast" and a "slow" power loop which obtain their error signal from the two single-element photo-diodes PDA1, PDA2 and actuate the power throughput of the modulator and the laser current, respectively. The "fast" loop compensates differential and the "slow" loop common mode power fluctuations.

Characterization and calibration of the OMS laser loops and their impact on performance is not the declared objective of this article and the interested reader is referred to dedicated publications with primary focus on the loop performance [8, 10]. For the measurements discussed in this article only the fast power loop, balancing and stabilizing the power ratio between the measurement and the reference beam, is relevant and the other loops have been left open.

## III. Phase-shifts and Channel noise

In the first step we aim to measure the phase-differences between the various processing channels and to characterize the phase-fluctuations. As a processing channel we understand the collective of photo-diode, cables from/to the phase-meter, the phase-meter itself (including input filters) and the processing software.

The relative phases of the interference signals coming from the photo-diodes in the four interferometers are generally arbitrary as they depend on a variety of things which we cannot precisely measure or control, such as exact position and orientation of test-mass 1 and 2 or minute differences in the optical path-length of the two beams. It is therefore not possible to extract readily exploitable channel calibration and noise data when operating the interferometers in nominal configuration as phase-sensitive detectors. An easy way to resolve this problem is to amplitude-modulate one of the beams at 1 kHz and switch the other beam off. The phase detected on each photo-diode is then given by the phase of the amplitude modulation alone which is the same on all diodes (neglecting time delays on the order of 1-2 ns due to different arm-lengths from FIOS output to photo-diode, which amount to phase-offsets of $2\pi \times 1 kHz \times 1 ns \sim 6 \times 10^{-6}\ rad$). As the OMS laser modulator does not support amplitude-modulation of the laser beams at 1 kHz (<50 Hz is supported), we have recourse to an auxiliary modulation bench. A commercial signal generator was frequency-locked to the DMU clock signal. The generator is configured to 1 kHz sine output which is fed to the amplitude control of the auxiliary modulation bench. The modulated laser beam is fed through the reference beam fibre injector onto the optical bench. The phase-signals are detected by the diodes, processed by the phase-meter and application software and recorded for a period of approximately 3 minutes, giving a total of 1800 points per channel.

When plotting the phase of any channel we observe a strong linear drift of ~8 rad/min and a much weaker quadratic drift of ~0.01 rad/min^2 common mode on all channels, which we attribute to an offset and drift, respectively, of the generator frequency with respect to the DMU master clock. In addition to this drift there is a common mode noise pattern on all channels which we also attribute to the signal source. In order to proceed with our investigation it is therefore necessary to subtract the phase of one channel -we choose quadrant A of diode PD1A and term it the "reference phase"- from all the others to cancel the common mode drift and the common mode noise related to the signal source. The resulting relative phases of all channels, all relating to quadrant A of PD1A as their reference, display a Gaussian sampling distribution of width $\sigma$ and centered around $\varphi_0$. As an example, the distributions of channels Q3 and Q4 of PD12A are displayed in Figure 2. The two distributions are offset by approximately -4 mrad and +1 mrad from the phase of channel (Q1, PD1A), respectively and have a Gaussian width (rms) of 1.61E-4 rad and 1.42E-4 rad respectively.

Analyzing the data for all channels, we find that the relative phases range from -8 mrad to +11 mrad and the distribution widths from 1.3E-4 to 1.6E-4 rad. Phase offsets between channels introduce a bias in the attitude measurements where a differential phase between diode quadrants is calculated (More details on differential phase measurements are found in the next section.). Our measurement data indicate that these phase-offsets are relatively small so that a maximal bias error of $10\,\mu rad$ in the attitude is introduced if we do not compensate them. However, to minimize bias errors we subtract the offsets by application of specific rotation matrices to the real and imaginary components of the complex amplitude vector of each channel. This process, which is executed automatically in the application software, effectively shifts the channel phase by the rotation angle $\theta$ and brings all phases "into alignment".

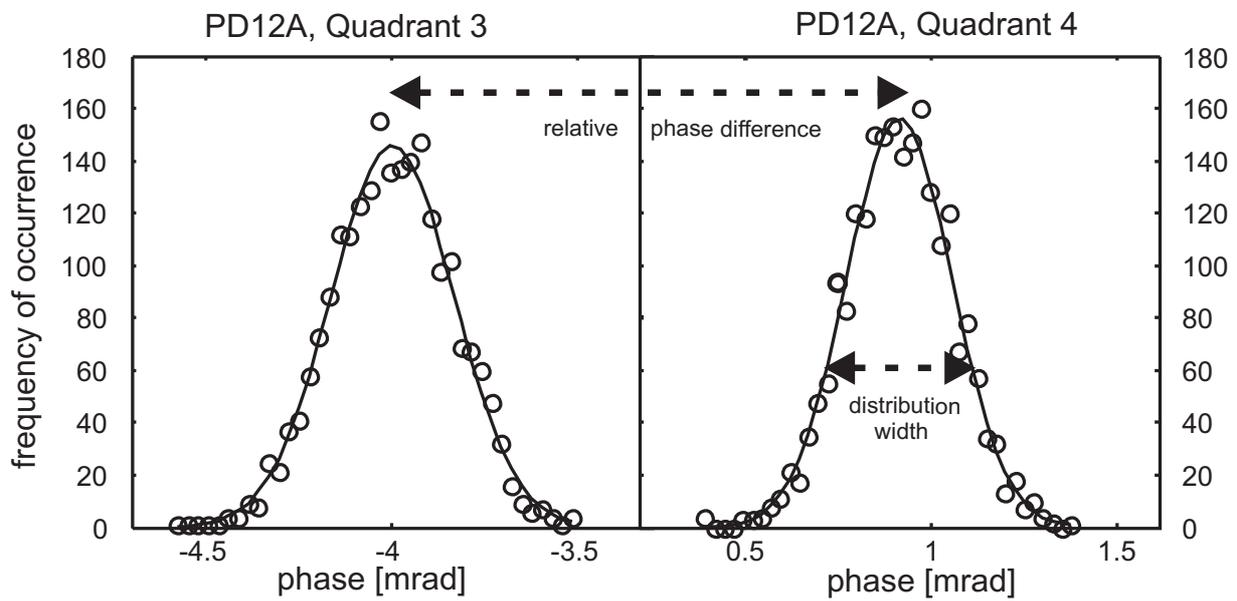

**Figure 2: The Gaussian distribution of relative (to Q1 of PD1A) phases for the processing channels Q3 and Q4 of diode PD12A.**

The fact that the distributions for the relative phase fluctuations are of similar width together with the assumption that the noise sources of any two channels do not correlate, imply that the distribution width of a single channel is a factor of $\sqrt{2}$ smaller than the relative distribution width, i.e. $\sigma_{relative} = \sqrt{\sigma_{channel}^2 + \sigma_{reference}^2} \approx \sqrt{2}\sigma_{channel}$.

This can be proven by showing that there is no (or only negligible) correlation between any two channels from the matrix of correlation coefficients. Unfortunately the correlation between two channels cannot be directly calculated for the respective channel phases $\varphi_i$ but only for the relative phases $\varphi_i - \varphi_0$ in order to remove the aforementioned common mode drifts and the noise of the signal source. Subtracting the reference phase $\varphi_0$ from the channel phase $\varphi_i$ then yields the relative channel noise $Z_i$:

$$\begin{aligned}\varphi_1 - \varphi_0 = Z_1 = X_1 - X_0 \\ \varphi_2 - \varphi_0 = Z_2 = X_2 - X_0\end{aligned}, \qquad (2)$$

where $X_i$ denotes the noise of channel 'i' and and $X_0$ the noise of the reference channel (quadrant A of diode PD1A).

The intra-channel correlation coefficients $r(i,j)$ are defined through the co-variances $C(Z_i, Z_j)$ as follows:

$$r(i,j) = \frac{C(Z_i, Z_j)}{\sqrt{C(Z_i, Z_i) C(Z_j, Z_j)}} = \frac{\langle (Z_i - \langle Z_i \rangle) \cdot (Z_j - \langle Z_j \rangle) \rangle}{\sqrt{\langle (Z_i - \langle Z_i \rangle)^2 \rangle \cdot \langle (Z_j - \langle Z_j \rangle)^2 \rangle}}, \qquad (3)$$

If we assume that the channels do not correlate, i.e. $C(X_i, X_j) = 0$, we find for r(i,j):

$$r(i,j) = \frac{\sigma_0^2}{\sqrt{(\sigma_1^2 + \sigma_0^2) + (\sigma_2^2 + \sigma_0^2)}} \approx \frac{1}{2} \qquad (4)$$

If, on the other hand, we assume there is correlation between any two channels we find that $r(i,j) = 0$ for negative correlation, i.e. $C(X_i, X_j) = -1$, and $r(i,j) = 1$ for positive correlation, i.e. $C(X_i, X_j) = 1$. As an example of the statistical analysis, the correlation matrix for the relative phases of the four channels of PDRA is given in table 1.

**Table 1: The correlation matrix between the relative phases of the quadrants of PDRA**

|       | $Z_1$ | $Z_2$ | $Z_3$ | $Z_4$ |
|-------|-------|-------|-------|-------|
| $Z_1$ | 1.00  | 0.45  | 0.49  | 0.48  |
| $Z_2$ | 0.45  | 1.00  | 0.46  | 0.43  |
| $Z_3$ | 0.49  | 0.46  | 1.00  | 0.47  |
| $Z_4$ | 0.48  | 0.43  | 0.47  | 1.00  |

We observe that all cross correlations are close to 0.5, indicating that there is no correlation between the noise sources of any two channels. The minor deviation from the exact value of 0.5 is explained by the fact that the noise distributions do not have the exactly same width, as we have already pointed out. At this point we have shown that the sensor processing chain is free of intra-channel noise correlations which could originate from cross-talk between diode quadrants, input filters or phase-meter channels with possibly serious impact on performance.

A further indication that the noise sources do not correlate is given by the combined longitudinal tracking phase $\psi_1$ which is calculated from the average phase of the four quadrants of PD1A minus the average reference phase of diode PDRA. Assuming the individual noise sources do not correlate, the combined noise floor $\sigma_{long}$ of the longitudinal tracking phase is then expected to be

$$\sigma_{long} \approx \frac{1}{4}\sqrt{\sum_{i=1}^{4}\overline{\sigma}^2} = \frac{\overline{\sigma}}{2} = \frac{1.4 \cdot 10^{-4}}{2} \, rad \ , \tag{5}$$

where $\bar{\sigma}$ denotes the average distribution width of the relative phase between two channels, as introduced above. The validity of the assumption made in the derivation of Equation 5 is confirmed by the measurement data.

At this point we would like to remind the reader that one major purpose of analyzing the channel noise is to find an estimate for the theoretical limit we can achieve in performance measurements of the longitudinal test-mass displacement. We therefore aim to scale the result of the root-mean-square noise of Equation 5 in such a way that it can be compared to performance measurements at nominal system configuration which were executed at a later point in time [10]. In order to do that we extract the utilization of ADC dynamic range from the measurement data and find that the peak-to-peak amplitude of the modulation is only 1/3 of the amplitude at nominal configuration (We had to restrict ourselves to small amplitudes in order to remain in the linear range of the beam amplitude modulator.) Theoretical analysis shows that in our operating range the effective channel noise scales inversely proportional to the utilization of dynamic range [14] so that the expected noise limit at nominal operation is 1/3 of the limit given in Equation 5.

Collecting all relevant factors and considering that the phase-data were output at 10 Hz, we arrive at $7\,\mu rad/\sqrt{Hz}$ for the linear spectral density of the longitudinal phase noise ($LSD_\psi$) which, upon application of the coupling factor $K_{long} \approx \lambda/4\pi$, translates into $0.62\,pm/\sqrt{Hz}$ for the displacement noise. Similarly, we obtain $10\,\mu rad/\sqrt{Hz}$ for the linear spectral density of the angular phase noise ($LSD_{DWS}$) which, upon application of the coupling factor $K^{-1}_{DWS} \approx 1/5000$, translates into $2\,nrad/\sqrt{Hz}$ for the attitude noise. The coupling factors for test-mass rotations ($K_{DWS}$) and test-mass translations ($K_{long}$) together with the definition of differential phase signals (DWS) will be discussed in the following sections. Note that the angular phase is calculated differently to the longitudinal phase so that the noise of the former scales as $\bar{\sigma}/\sqrt{2}$ whereas the noise of the latter scales as $\bar{\sigma}/2$ (see Equation 5). We would like to remind the reader that although the noise limits were derived from data taken during only 180 seconds of measurements, the corresponding noise floor is applicable for the entire frequency spectrum- assuming the processing channels have reached a quasi-stationary state. Equation 6 summarizes the result for the lower noise threshold applicable to performance measurements of the test-mass position:

$$
\begin{aligned}
LSD_\psi &= \frac{1.4 \cdot 10^{-4}\, rad}{\sqrt{10Hz}} \frac{1}{2}\frac{1}{3} = 7\times 10^{-6}\, rad \cdot Hz^{-1/2} \xrightarrow{\times \lambda/4\pi} 0.62\, pm\, Hz^{-1/2} \\
LSD_{DWS} &= \frac{1.4 \cdot 10^{-4}\, rad}{\sqrt{10Hz}} \frac{1}{\sqrt{2}}\frac{1}{3} = 1\times 10^{-5}\, rad \cdot Hz^{-1/2} \xrightarrow{\times K_{DWS}^{-1}} 2.0\, nrad\, Hz^{-1/2}
\end{aligned}
\qquad (6)
$$

These values constitute lower limits for the noise (best possible performance). The actual performance is generally expected to be lower due to laser frequency fluctuations and variations in optical path-length which contribute significantly to the total noise level. However, once the laser frequency and OPD stabilization loops operate optimally, we should be able to approach the noise levels given in Equation 6. In order to achieve the primary mission goal, i.e. the sensitivity in the overall measurement of residual acceleration as given in Equation 1 of the introduction, the total noise level of the OMS measurements within the measurement band-width (3 mHz to 30 mHz) is required to be lower than $6.4\, pm/Hz^{1/2}$ for the longitudinal displacement between the two-test-masses and 10 $nrad/Hz^{1/2}$ for the test-mass attitude. We conclude that the noise from phase-meter, electronics, photo-diodes and digital processing is compliant with the requirement (on a measurement timescale of ~180 s), leaving laser frequency fluctuations and OPD noise as the major remaining noise sources.

## IV. Interferometer coupling parameters

*The interferometer signals and their reference frame*

The OMS application software on the data-management unit (DMU) receives the complex phase-data and processes them to obtain what is commonly referred to as DC- and DWS-alignment signals, where DWS stands for "differential wave-front sensing".

The DC-signals are calculated from the DC-values of the discrete Fourier transform. The DC-signal $DC_\phi$ for the horizontal angle $\phi$, which corresponds to a rotation around the z-axis perpendicular to the optical table (see Figure 3a), is defined as the normalized difference in laser power between the left and right diode half:

$$DC_\phi = \frac{DC_A + DC_C - DC_B - DC_D}{DC_A + DC_B + DC_C + DC_D} \tag{7}$$

Similarly, the DC-signals $DC_\eta$ for the vertical angle $\eta$, which corresponds to a rotation around the y-axis lying in the optical table and perpendicular to the x-axis connecting the two test masses, are defined as the normalized difference in laser power between the upper and lower diode half. An illustration of the applicable coordinate system and the naming convention of the diode quadrants is given in Figures 3 (a) and (b), respectively.

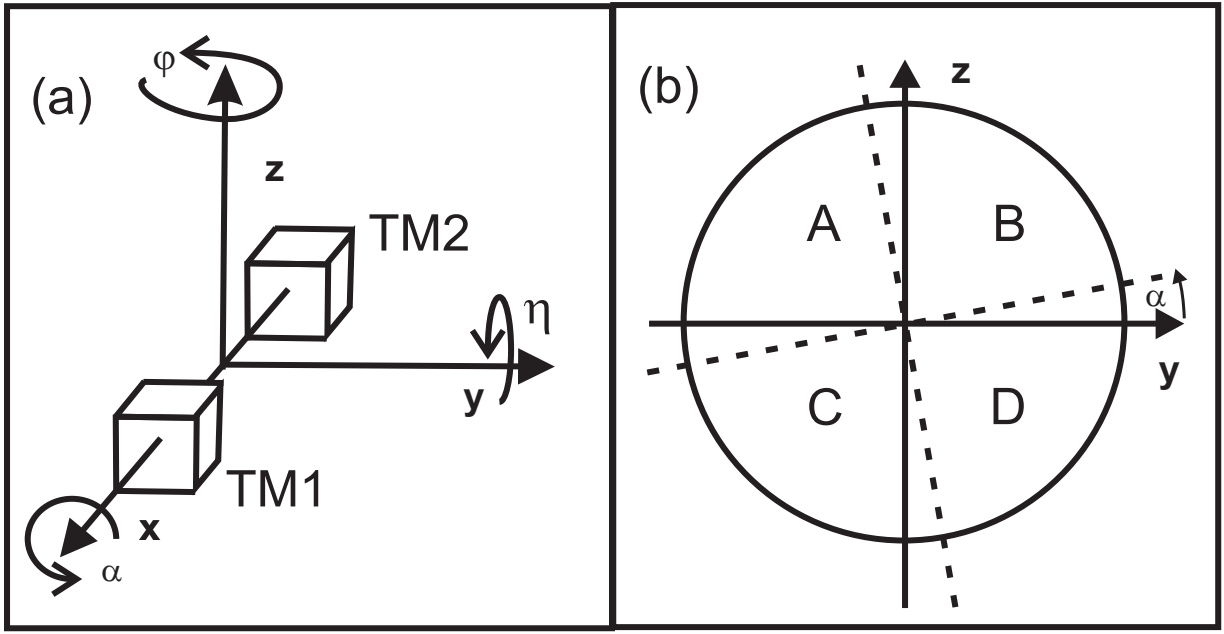

**Figure 3: (a) Definition of the optical bench reference frame. (b) The diode quadrants are labeled A,B,C,D from top to bottom. The PZT sweep axes (dotted lines) are rotated with respect to the quadrant diode axes (solid lines).**

The DWS-signals are calculated from the complex amplitude of the Fourier transform. After some rescaling of the complex amplitude and applying the rotation matrices discussed in the previous section we obtain the phasor $F$. The "horizontal" DWS-signal is defined as the phase difference between the left and the right diode half, the "vertical" DWS-signals as the phase-difference between upper and lower diode half:

$$DWS_\phi = \arg\left(\frac{F_A + F_C}{F_B + F_D}\right) \qquad DWS_\eta = \arg\left(\frac{F_A + F_B}{F_C + F_D}\right) \qquad (8)$$

Whereas the DWS-signals are very sensitive and allow highly accurate measurements at small angles ($< 200 \mu rad$), the DC-signals are much less sensitive but allow to measure angles over a much larger range ($< 2000\,\mu rad$), which finds its main application in the initial coarse alignment of the test-masses.

Assuming a linear relationship between the test-mass attitude degrees of freedom and the interferometer DWS- and DC-signals, and referring to the basic operating principle of the interferometers x1 and x1-x2 as shown in Figure 1, we define the interferometer coupling constants K1-K6 and K11-K16 through the following set of equations:

$$\begin{aligned}
DWS_1^\phi &= K_1 \phi_1 & DC_1^\phi &= K_{11} \phi_1 \\
DWS_1^\eta &= K_2 \eta_2 & DC_1^\eta &= K_{12} \eta_2 \\
DWS_{12}^\phi &= K_3 \phi_1 + K_4 \phi_2 & DC_{12}^\phi &= K_{13} \phi_1 + K_{14} \phi_2 \\
DWS_{12}^\eta &= K_5 \eta_1 + K_6 \eta_2 & DC_{12}^\eta &= K_{15} \eta_1 + K_{16} \eta_2
\end{aligned} \qquad (9)$$

where indices "1" and "12" for the DWS- and DC-signals refer to interferometers x1 and x1-x2, respectively, and the indices "1" and "2" for the test-mass angles $\phi, \eta$ refer to test-mass 1 and 2, respectively.

*Theoretical derivation of coupling constants*

From the definition in Equation 7 it is easy to derive an analytical expression for the DC-signals as a function of the beam displacement from the quadrant diode center: Consider a Gaussian measurement and reference beam with intensity $I_M(y,z), I_R(y,z)$, center along the y-direction $y_{0M}, y_{0R}$, beam waist $w_M, w_R$, and total power $P_M, P_R$, respectively. We then find for the horizontal DC-signal

$$DC_\phi = \frac{\int\limits_{-\infty}^{\infty} dz \int\limits_{-\infty}^{0} dy \left(I_M(y,z) + I_R(y,z)\right) - \int\limits_{-\infty}^{\infty} dz \int\limits_{0}^{\infty} dy \left(I_M(y,z) + I_R(y,z)\right)}{\int\limits_{-\infty}^{\infty} dz \int\limits_{-\infty}^{\infty} dy \left(I_M(y,z) + I_R(y,z)\right)} \quad (10)$$

$$= \frac{P_M}{P_M + P_R} erf\left(\frac{\sqrt{2} y_{0M}}{w_M}\right) + \frac{P_R}{P_M + P_R} erf\left(\frac{\sqrt{2} y_{0R}}{w_R}\right)$$

When the dummy-mirror is tilted by an angle $\varphi$, the measurement beam center moves accordingly by a distance of $y_{0M} = 2\varphi L_{TM}$, where $L_{TM}$ is the lever arm from test-mass to diode, whereas the reference beam remains static. Substituting the expression for the beam displacement $y_{0M}$ into Equation 10 and expanding it to first order we obtain an expression for the DC-coefficients

$$K_{DC} = \frac{P_M}{P_M + P_R} \sqrt{\frac{2}{\pi}} \frac{4 L_{TM}}{w_M} \quad (11)$$

Note that the coefficient depends on the beam power ratio, the lever arm length and the beam waist. We give an example for the expected order of magnitude for the DC-coefficient: Assuming that the beam powers are equal and that the beam waist is 1 *mm*, and considering that the lever arm length from dummy-mirror 1 to PD1A is $L_{TM} = 29.5\,cm$, we obtain $K_{11} = 470$.

Calculation of the DWS-coupling coefficients is more difficult as the coefficients depend strongly on the wavefront curvatures of the two interfering beams. Assuming the curvatures are small, a simplified expression is found [15] for the DWS-signal and its linearized slope, the $K_{DWS}$ coefficient:

$$DWS_\varphi \approx a\tan\left(erfi\left(\frac{2\pi}{\lambda}\frac{w_M}{\sqrt{2}}2\varphi\left(1-\frac{L_{TM}}{R}\right)\right)\right),$$

$$K_{DWS} \approx 4\sqrt{2\pi}\frac{w_M}{\lambda}\left(1-\frac{L_{TM}}{R}\right)$$

(12)

where "erfi" is the imaginary error-function defined by $erfi(z) = -i \cdot erf(z)$ and $R$ is the beam radius at the interference point. It is important to note that the coefficient depends on beam waist, lever arm length and wave-front curvature. As an example for the expected order of magnitude, we consider the interference on PD12A: Assuming that the beam waist 1mm, the beam radius of curvature is $R = 1.4\,m$, and the lever-arm length $L_{TM} = 52.2\,cm$, we find $K_3 = 5911$.

## *Measurement approach for the coupling constants*

In order to find the accurate values of the K-coefficients we determine the linear dependence of the DC- and DWS-signals on the tilt angle of the test-mass (represented by a dummy mirror). The mirror is attached to the surface of a 3-axis piezoelectric transducer (PZT) to accomplish the tilt. The PZT device consists of a metal cylinder containing three identical PZTs which are symmetrically placed around the central axis. A suitable PZT driver applies variable voltages to the PZTs which affects a corresponding tilt across an axis determined by the voltage ratios. Through appropriate choice of two "orthogonal" sets of basis voltages, the front-face mirror can be tilted across either of two corresponding orthogonal directions. In particular, it can be tilted horizontally (angle $\phi$) or vertically (angle $\eta$). We could adjust the basis voltages up to a certain accuracy so as to make the tilt axes orthogonal to within 1.5 degrees. Due to limitations of the test-setup the axes of the PZT assembly cannot be aligned with the reference axes of the bench in a well controlled way. This results in an -a priori unknown- misalignment of the PZT tilt axes with respect to the optical bench frame by an angle $\gamma$ on the order of 3 degrees. All measurements rely on signals from the quadrant-diodes, which in turn have an unknown tilt of their quadrant axes with respect to the optical bench frame. We can therefore only determine the angle $\alpha$ between the PZT tilt axes and the diode quadrant axes, as shown in Figure 3b, but not the angle $\gamma$.

The calibration coefficients K are defined as the slope of the DWS/DC-signal against mirror tilt angle in the linear central region around the angle $\varphi = 0$, as described in Equations 11 and 12. In our measurements we determine the slope of DWS/DC-signal against applied PZT driver voltage. We therefore have to divide this slope by a "PZT calibration factor", describing the linear relation (valid for small tilt angles) between mirror tilt angle and the applied PZT voltage, to obtain the required coefficient.

These measurements are either performed "point-by-point", where the driver voltage is stepwise incremented, or by application of a sinusoidal voltage to the PZT driver. In the latter case, the amplitudes of the sinusoidal response in the DWS/DC signals are determined instead of fitting the linear central slope. The amplitudes are then divided by a different PZT calibration factor, which relates amplitude of the dummy mirror tilt angle to the amplitude of the sinusoidal voltage applied to the driver.

Figure 4 displays a summary of the point-by-point calibration measurements for a horizontal tilt of dummy-mirror 1. In Figure 4a the DC-signals are plotted for the case where the dummy mirror is at first turned counter-clockwise and then clockwise. A PZT hysteresis effect is clearly visible so that the mirror tilt differs between the path where the PZT voltage is increased (counter-clockwise rotation, upper curve) and the return path (lower curve). However, the linear central region, critical for obtaining the coupling coefficients, yields the same slope to within $\pm 2\%$ for both curves which is of acceptable accuracy. The upper curve is also fitted by an error function, represented by the solid line, from which the width of the Gaussian beam in horizontal direction can be directly determined. Additionally, the calibration constant K11, as defined in Equation 11, is extracted from the linear region of the error-function and we find K11=510, in accordance with the theoretical expectations.

The contrasts of the interference pattern are displayed in Figure (4c) for the same counter-clockwise (right curve) and clockwise (left curve) sweeps as in (4a). The contrasts grow and peak in the region where the DC- and DWS signals display a linear dependence on the test-mass tilt. Note that the curves for the contrasts have a Gaussian shape in accordance with the Gaussian beam profile.

Figure (4b) displays the DWS-signals which for the same counter-clockwise (upper curve) and clockwise (lower curve) rotations as in Figure (4a). The DWS-signals show a similar error-function-like dependency on the dummy mirror tilt angle as the DC-signals, albeit in much smaller range between -1 mrad and +1 mrad. The linear central region of the DWS-signal curves is approximately 500 micro-rads in width and its slope determines the coefficient K1 of Equation 12. The fit to this linear region, which also corresponds to the 5 points with maximum contrast in the right curve of Figure 4b, yields K1=5190, in accordance with the theoretical expectations.

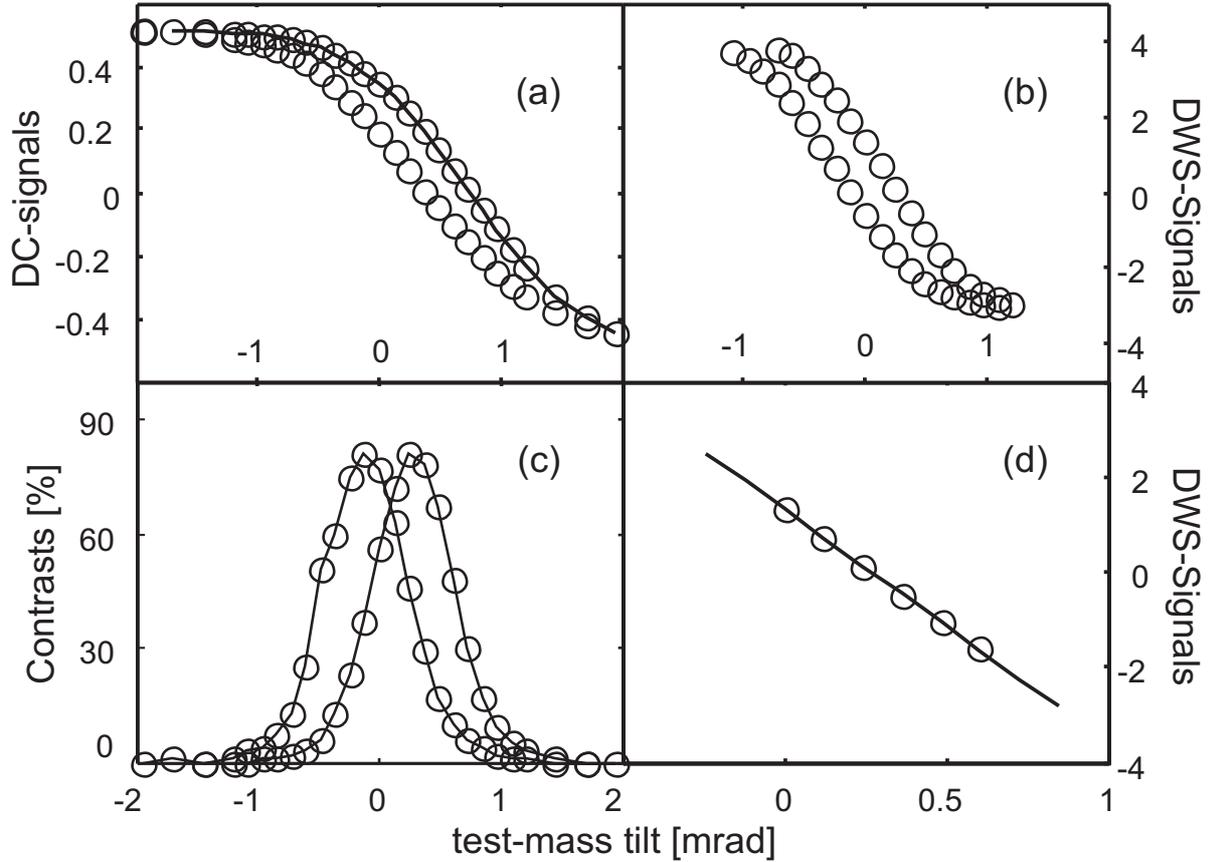

**Figure 4: All plots refer to the same measurement series and all data were recorded simultaneously. (a) The DC-signals are plotted for a counter-clockwise (upper curve) and clockwise (lower curve) rotation of the dummy mirror. The upper curve is fitted by an error function (solid line). (b) The DWS-signals for counter-clockwise (upper curve) and clockwise (lower curve) rotation of the dummy mirror. (c) The interference contrasts for counter-clockwise (right curve) and clockwise (left curve) rotation. The points are interconnected for illustration only. (d) The linear central region of the upper curve in (b) was fitted to extract the calibration coefficient K1.**

*Discussion of measurement results*

In a similar way to the measurements described in the preceding section we obtained all the interferometer coupling coefficients which are listed in table 2. The estimated total error of the K-coefficients is 5% of the absolute value. It is given by a combination of fitting error and systematic errors due to PZT hysteresis and non-linearity.

**Table 2: The measured coupling-coefficients, K1-K6 (DWS), K11-K16 (DC).**

| K1 | K2 | K3 | K4 | K5 | K6 | K11 | K12 | K13 | K14 | K15 | K16 |
|---|---|---|---|---|---|---|---|---|---|---|---|
| 5190 | 4963 | 5174 | 7281 | 4898 | 6793 | 510 | 595 | 615 | 191 | 718 | 228 |

The PZT non-linearity is especially problematic for increasingly large tilt-angles which require high bias voltages. To derive the K-coefficients in the linear central region this is unproblematic in contrast to a full sweep of the beam across the diode, as required to fit the error-function of the beam profile to determine the width. In that case the PZT non-linearity occurring towards the extremes of the sweep leads to an underestimate of the beam width. However, a better estimate of the beam width can be found when solving Equation 11 for $w_M$ and inserting the measured value $K_{DC}$, the known arm-length $L_{TM}$, and the measured Power ratio of the beams. The latter is found from the fit of the whole PZT sweep with an error-function, whose amplitude, according to Equation 11, is given by the beam power ratio. When comparing the beam-width found from the coupling coefficient to the beam width found from the error-function fit we find that the latter has been consistently underestimated by ~5% in all cases. Taking this into account, the adjusted horizontal and vertical widths of the measurement beam are found to be (9.37E-4 m, 7.73E-4 m) on PD1A, and (11.02E-4 m, 9.44E-4 m) on PD12A, respectively. This clearly indicates that the beam is elliptic and not circular (ellipticity ~1.20). The implications of ellipticity immediately become evident in the difference between "vertical" and "horizontal" coupling parameters of table 2. Further consequences are discussed in the next section.

It is possible to deduce the radius of beam curvature of the measurement beam on PD12A from the K-coefficients and therefore completely determine the Gaussian beam parameters [15]. We find R~1.40 m for the measurement beam on PD12A. We also obtain the ratio of beam powers from the DC-signals recorded during a long-range PZT sweep and find that the two interfering beams have exactly the same power on PD1A but $P_R = (1.47 \pm 0.04) \cdot P_M$ on PD12A. The most likely explanation is that the measurement beam is split 3:2 at BS3 (see Figure 1) which results in lower beam power in the measurement arm to TM2 and therefore unbalanced beams at PD12A. Note that power measurements at any location in the interferometers are precluded by the lack of space to insert a separate power sensor and generally by the stringent requirements on handling, cleanliness, and contact avoidance.

## V. Interferometer Cross-coupling terms

Cross-coupling between the two tilt axes, i.e. the appearance of a "false" signal along one tilt-direction although the test-mass tilts along the orthogonal direction, is an undesirable effect that frequently becomes apparent. When we modulate the dummy mirror tilt in the horizontal direction, we observe that the primary horizontal $DC_\varphi / DWS_\varphi$-signals are accompanied by a tiny residual oscillation in the vertical direction. The origin of this residual oscillation is the imperfect alignment between the mirror tilt axes and the diode quadrant axes, as mentioned before and illustrated in Figure 3b. The relative orientation of the two axes pairs is easily inferred from the ratio of the two oscillation amplitudes.

From certain measurements, where data were simultaneously recorded on interferometers x1 and x1-x2, we can additionally determine the relative angle between the diode quadrant reference frames of PD1A and PD12A. We find that there is an angle of approximately 3 degrees between PD1A and PD12A which is also an indication for the degree of accuracy with which the diode quadrant frame has been aligned with the reference frame of the optical bench. We shall now investigate the impact of such a misalignment on the cross-coupling between directional degrees of freedom and the resulting steady states in a closed feedback system as used in the mission when the test-masses are actually floating [5, 6, 7]. In the following derivation we revert to the defining Equations (9) of the coupling coefficients for the DC-signals (for DWS-signals an analogous derivation applies) and introduce the following equations for ease of notation:

$$\begin{pmatrix} \underline{DC}_1 \\ \underline{DC}_2 \end{pmatrix} = \begin{pmatrix} A & 0 \\ B & C \end{pmatrix} \begin{pmatrix} \underline{\varphi}_1 \\ \underline{\varphi}_2 \end{pmatrix}$$
$$\underline{\varphi}_i = \begin{pmatrix} \phi_i \\ \eta_i \end{pmatrix}, \underline{DC}_i = \begin{pmatrix} DC_i^\phi \\ DC_i^\eta \end{pmatrix}; A = \begin{pmatrix} K_{11} & 0 \\ 0 & K_{12} \end{pmatrix}; B = \begin{pmatrix} K_{13} & 0 \\ 0 & K_{15} \end{pmatrix}; C = \begin{pmatrix} K_{14} & 0 \\ 0 & K_{16} \end{pmatrix}$$
(13)

## *The primary reference frames and their symmetries*

We shall at first only look at the equations governing TM1. Note that the coupling constants $K_{11}, K_{12}$ are not identical but differ by 10%-20%. As discussed in the previous section, their difference in value originates from anisotropies in the beam parameters. In our discussion it is useful to remember that there are essentially three different reference frames:

1. **The optical bench frame:** Measurement (and per default reference) beam are nearly perfectly aligned with this frame when the DWS-signals are zero.
2. **The diode quadrant frame:** The diodes are our primary sensor, all processing is based on their signals and all output (DC- and DWS-signals) is therefore referenced to the quadrant frame.

3. **The beam frame:** The ellipsoidal beam shape (and its associated beam curvature radii) defines an intrinsic reference frame through its major and minor axes, which are not generally aligned with either optical bench or quadrant reference frames.

*The effect of beam asymmetry on the coupling coefficients*

We shall first examine the impact of a rotated beam frame. We assume that the coupling constants $K_{11}, K_{12}$ were initially measured with both, quadrant and beam frame, aligned with another. We then rotate the beam axes counter-clockwise by an angle $\beta$ with respect to the diode quadrants. After some rather lengthy calculations, following a similar ansatz to the one of Equation 10, we find for the effective K-coefficients $K_{11}^{eff}(\beta), K_{12}^{eff}(\beta)$ in the quadrant frame:

$$\begin{aligned} K_{11}^{eff} &= \left(K_{11}^{-2}\cos^2\beta + K_{12}^{-2}\sin^2\beta\right)^{-1/2} \\ K_{12}^{eff} &= \left(K_{12}^{-2}\cos^2\beta + K_{11}^{-2}\sin^2\beta\right)^{-1/2} \end{aligned} \qquad (14)$$

We observe that the coupling coefficients are changing with increasing rotation angle of the ellipse so that gradually $K_{11}$ turns into $K_{12}$ and vice versa. Note that there are no off-diagonal coupling elements introduced by rotation of the beam ellipsoid with respect to the quadrant frame but the value of the diagonal elements is changing accordingly.

*The effect of photo-diode misalignment on the coupling coefficients*

We shall now investigate the impact of misalignment between bench frame (associated parameters have a tilde) and the quadrant frame. We assume that the bench frame is rotated clockwise by an angle $\alpha_1$ with respect to the quadrant frame of PD1A. The tilt angles $\underline{\varphi}_1$ in the quadrant frame are obtained from the tilt angles $\underline{\tilde{\varphi}}_1$ in the optical bench frame through an orthogonal transformation, represented by the matrix $R(\alpha_1)$. This is depicted in Figure 5a.

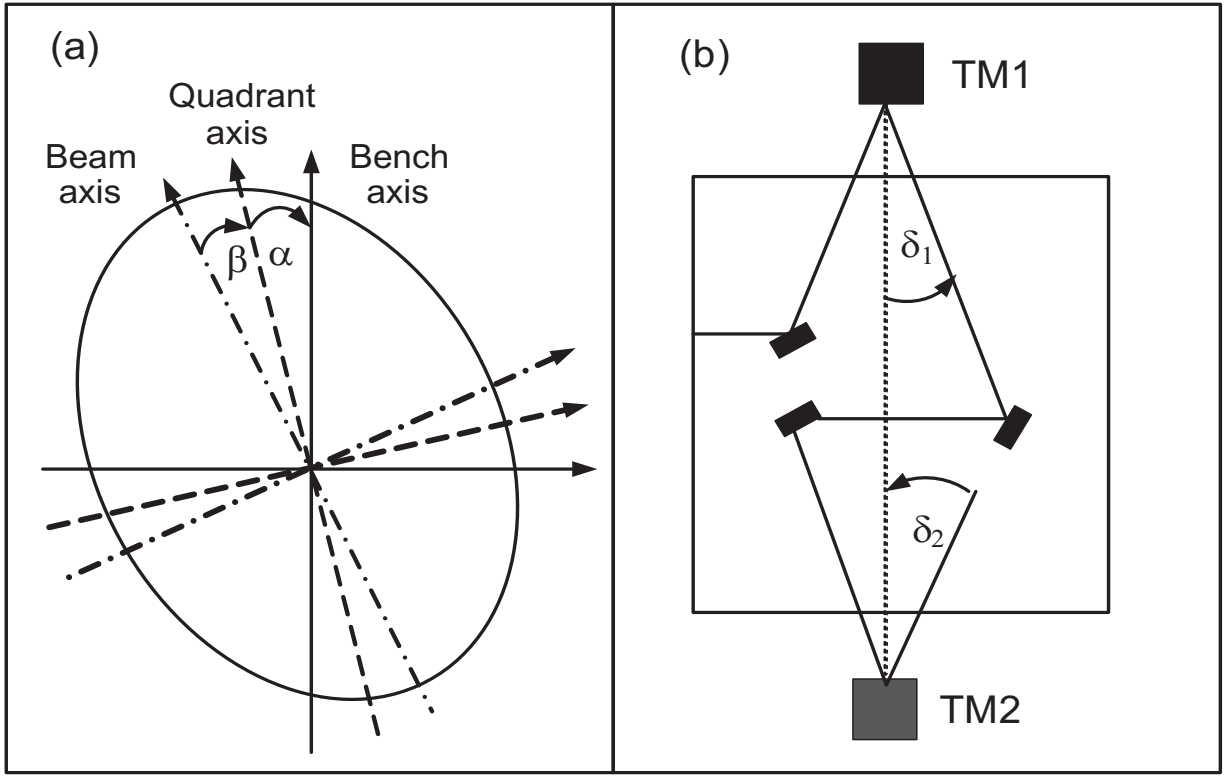

**Figure 5: (a) The three main reference frames are depicted. The quadrant reference frame is rotated by an angle $\beta$ with respect to the beam frame. The optical bench frame is rotated by an angle $\alpha$ with respect to the quadrant frame. (b) A simplified schematics of the measurement beam path. The incidence angles $\delta$ on test-mass 1 and test-mass 2 are designed to be equal under nominal conditions.**

Considering $\underline{\tilde{\varphi}}_1 = R(\alpha_1) \cdot \underline{\varphi}_1$ we obtain from Equation 13:

$$\begin{aligned}\underline{DC}_1 = A \cdot \underline{\varphi}_1 = A \cdot R(\alpha_1) \cdot \underline{\tilde{\varphi}}_1 = \begin{pmatrix} K_{11}\cos\alpha_1 & K_{11}\sin\alpha_1 \\ -K_{12}\sin\alpha_1 & K_{12}\cos\alpha_1 \end{pmatrix} \cdot \begin{pmatrix} \tilde{\phi}_1 \\ \tilde{\eta}_1 \end{pmatrix} \\ \neq R(\alpha_1) \cdot A \cdot \underline{\tilde{\varphi}}_1 \end{aligned} \quad (15).$$

Note that the rotation matrix $R(\alpha_1)$ and the coupling matrix $A$ do not commute unless the rotation angle is zero ($\alpha_1 = 0$) or the beam symmetry is circular and not elliptic ($K_{11} = K_{12}$). If the two commute we find that the DC-signals in the new reference frame are simply rotated the same way as the test-mass angles.

Equation 15 implies that there is always residual coupling into the vertical direction when the test-mass is moved horizontally (in the optical bench frame) and vice versa. However, this does not really change the steady state of the test-mass which is directed towards DC=0 (analogously for DWS=0) by the drag-free attitude-control system (DFACS). The error in the DC-control signal, which is induced by a (small) misalignment between quadrant and bench frame, is proportional to the test-mass angles $\tilde{\varphi}_1$ and therefore vanishes when the test-mass approaches DC=0 (analogously for DWS), i.e. $\tilde{\varphi}_1 \to 0$. The error as a percentage of the overall signal for a misalignment of $\alpha_1 = 3\,\text{deg}$ is given by $e \approx \alpha_1 = 0.05$, i.e. it is on the order of 5%.

Proceeding in a similar way as for interferometer x1 and considering a relative rotation by the angle $\alpha_2$ between the quadrant frame of diode PD12A and the optical bench frame, we find an expression for the DC-signals in interferometer x1-x2 from Equation 13:

$$\underline{DC}_2 = B \cdot \underline{\varphi}_1 + C \cdot \underline{\varphi}_2 = B \cdot R(\alpha_2) \cdot \underline{\tilde{\varphi}}_1 + C \cdot R(\alpha_2) \cdot \underline{\tilde{\varphi}}_2 \tag{16}$$

Equations 15 and 16 can be solved for the test-mass angles in the bench frame, which are the actual quantities fed back to DFACS

$$\begin{aligned} \underline{\tilde{\varphi}}_1 &= R^{-1}(\alpha_1) \cdot A^{-1} \cdot \underline{DC}_1 \\ \underline{\tilde{\varphi}}_2 &= R^{-1}(\alpha_2) \cdot C^{-1} \left( \underline{DC}_2 - B \cdot R(\alpha_2 - \alpha_1) \cdot A^{-1} \cdot \underline{DC}_1 \right) \end{aligned} \tag{17}$$

We observe in Equation 17 that the test-mass angles in the optical bench frame are obtained from the test-mass angles in the quadrant frames through orthogonal transformations $R^{-1}(\alpha_1)$ and $R^{-1}(\alpha_2)$, respectively. However, there is one additional step: the angles of test-mass 1, $\underline{\varphi}_1 = A^{-1} \cdot \underline{DC}_1$, must be "adjusted" by the relative angle $\alpha_2 - \alpha_1$ between diode frames PD1a and PD12A through application of the orthogonal matrix $R(\alpha_2 - \alpha_1)$.

As long as the diode quadrant frame misalignments (relative to the optical bench as well as relative to one another) are sufficiently small, the couplings introduced by the orthogonal matrices $R(\alpha_1), R(\alpha_2), R(\alpha_1 - \alpha_2)$ are weak and these terms may be neglected. Even if $\alpha_1, \alpha_2$ were precisely known, the application software does not presently have the capability to compensate the rotations as described in Equation 17. However, this is not too problematic as we conclude that the test-mass angles converge towards DC=0 considering that the error introduced by neglecting the aforementioned couplings also converges to zero.

*Tracking accuracy of the longitudinal test-mass position*

We also test the ability of the interferometers x1 and x1-x2 to continuously track longitudinal movements of the dummy mirrors over distances of several hundred microns and investigate how well the individual movements of TM1 and TM2 de-couple from another.

The two interferometers record the average phase of all four quadrants of their respective diodes and subtract the reference phase $\Psi_R$ to obtain the "longitudinal phases" $\Psi_1$ and $\Psi_{12}$. These relate to the longitudinal displacement $d_x$ as follows:

$$d_x = \frac{\lambda}{4\pi \cos \delta} \psi, \qquad (17)$$

where $\delta = 4.5 \deg$ is the angle at which the beam is incident on the test-mass at nominal configuration. We find that the two interferometers track perfectly well a sinusoidal motion of dummy mirror 1 of amplitude ~100 microns. As the longitudinal phase $\Psi_{12}$ of interferometer x1-x2 is proportional to the relative displacement between TM1 and TM2, $\Psi_{12}$ should equal $\Psi_1$ (except for a constant term), if only dummy-mirror 1 is moved. However, this only holds if the two incidence angles $\delta_1$ and $\delta_2$ are exactly identical, otherwise the motion of TM1 and TM2 cannot be fully separated. The cross-coupling term $C_{long}$ is determined by the ratio of the two incidence angles: $C_{long} = (\delta_2 - \delta_1)/\delta_1$.

Subtracting $\Psi_1$ from $\Psi_{12}$, we find some residual noise but no visible remaining oscillation that could be an indication for cross-coupling. From the ratio of the standard deviation of the phase-difference to the standard deviation of the phase we find an upper threshold for the cross-coupling term $C_{long} < 3 \cdot 10^{-4}$ which implies $(\delta_2 - \delta_1) < 25 \, \mu rad$.

## VI. Conclusions and outlook

We successfully operated and investigated the complete engineering model of the Optical Metrology System for the LISA Pathfinder mission for the first time. We measured and analyzed the channel noise in detail and derived an upper limit for the expected system performance in the absence of laser frequency fluctuations. While cross-correlations between channel noise sources were shown to be negligible, the measured inter-channel phase-differences were successfully compensated.

The coupling constants relating test-mass attitude to differential phase and DC-signals were theoretically derived and compared to the measurements. The three principal interferometer frames (diode quadrant frame, optical bench frame, beam anisotropy frame) were introduced and the impact of a general misalignment between them was discussed. In addition to the capability of accurately determining the test-mass attitude we also demonstrated the capability of the system to track the test-mass position over long distances. The cross-coupling of signals describing test-mass 1 motion to signals describing test-mass 2 motion was investigated and an upper limit found.

The measurements described in this article refer to the engineering model of the optical metrology system. We already found compliance of the engineering model with all relevant system and mission requirements as far as applicable. The actual flight model is currently being built and should improve significantly on several deficiencies found in the engineering model, specifically on general noise characteristics, utilization of ADC dynamic range, beam isotropy, and relative misalignment between diode and optical bench frame.

## Acknowledgements

We gratefully acknowledge the financial support by Deutsches Zentrum für Luft- und Raumfahrt (DLR) to perform the experiments and measurements with the optical metrology system described herein on the premises of the Albert Einstein Institute. We would like to thank all personnel of the institute and of Astrium who contributed. G.H. would like to thank David Hoyland, University of Birmingham, U.K., for his extensive help and many enlightening "noise" discussions. G.H. is grateful to I.C.W of the R.B.C. for providing support and means to write this article.